# A d-Sequence based Recursive Random Number Generator

Abhishek Parakh
Louisiana State University

## Introduction

Binary d-sequences were investigated for their cryptographic properties by Subhash Kak in [1,2], who also examined their application to watermarking [3], and made a proposal for their use as random number generators (RNGs) [4] in a manner analogous to the iterative squaring done in the BBS method [5,6].

In this paper we propose a new recursive technique for the use of d-sequences to generate random numbers.

## Decimal Sequences

A decimal sequence is obtained when a number is represented in a decimal form in a base $r$ and it may terminate, repeat or be aperiodic. For a certain class of decimal sequences of $1/q$, $q$ prime, the digits spaced half a period apart add up to $r-1$, where $r$ is the base in which the sequence is expressed. Decimal sequences are known to have good cross- and auto-correlation properties and they can be used in applications involving pseudorandom sequences. The following section describes the properties of decimal sequences [1,2]:

**Theorem 1:** Any positive number x may be expressed as a decimal in the base *r*

$$A_1 A_2 \ldots A_{s+1} . a_1 a_2 \ldots \tag{1}$$

where $0 \le A_i < r$, $0 \le a_i < r$, not all $A$ and $a$ are zero, and an infinity of the $a_i$ are less then $(r-1)$. There exists a one-to-one correspondence between the numbers and the decimals and

$$x = A_1 r^s + A_2 r^{s-1} + \ldots + A_{s+1} + \frac{a_1}{r} + \frac{a_2}{r^2} + \ldots \tag{2}$$

That the decimal sequences of rational and irrational numbers may possibly be used to generate pseudo-random sequences is suggested by the following theorems of decimals of real numbers.

**Theorem 2:** Almost all decimals, in any base, contain all possible digits. The expression almost all implies that the property applies everywhere except to a set of measure zero.

**Theorem 3:** Almost all decimals, in any base, contain all possible sequences of any number of digits.

Theorems 2 and 3 guarantee that a decimal sequence missing any digit is exceptional.

The binary d-sequence is generated by means of the algorithm:

$$a(i) = 2^i \bmod p \bmod 2 \tag{3}$$

where $p$ is a prime number and $a(i) = 2^i \bmod p$ is a simple d-sequence. The maximum length $(p-1)$ sequences are generated when 2 is a primitive root of $p$. When the binary d-sequence is of maximum length, then bits in the second half of the period are the complements of those in the first half.

It is easy to generate d-sequences, which makes them attractive for many engineering applications.

It was shown in [2] that it is easy to find $i$ given $\log_2 p$ bits of $a(i)$, therefore, d-sequences cannot be directly used in random number generator applications.

## Kak RNG [4]

By adding together two or more different binary d-sequences (obtained using primes $p_1, p_2, ....$), one is able to introduce non-linearity in the generation process and the resulting sequence becomes a good candidate for use as random sequence.

$$a(i) = 2^i \bmod p_1 \bmod 2 \;\oplus\; 2^i \bmod p_2 \bmod 2 \;\oplus\; 2^i \bmod p_3 \bmod 2 \;\oplus\; ... \quad (4)$$

where $\oplus$ is the modular 2 addition and $p$ denotes a (ideally very large) prime number.

If the individual sequence is maximum length, then the period of the sum will be

$$lcm\{\,(p_1 - 1), (p_2 - 1), (p_3 - 1), ...\}$$

where $lcm(a,b)$ means the Least Common Multiple of $a$ and $b$. Since we are dealing with only positive integers, we say $c$ is a least common multiple of $a$ and $b$ if

1. $c$ is a multiple of $a$ and $b$;
2. any multiple of $a$ and $b$ is a multiple of $c$.

For randomly chosen primes we do not know if the starting number is a primitive root, therefore, the actual period would be a divisor of $lcm\{\,(p_1 - 1), (p_2 - 1), (p_3 - 1), ...\}$.

If we choose a seed S, which is relatively prime to each $p_i$, and the order of $S$ does not divide $(p_i - 1)$ for all $i$, then the power-exponent random number generates bits according to the following algorithm:

$$\begin{aligned} a(0) &= S \bmod p_1 \bmod 2 \oplus S \bmod p_2 \bmod 2 \\ a(1) &= S^2 \bmod p_1 \bmod 2 \oplus S^2 \bmod p_2 \bmod 2 \\ a(2) &= S^4 \bmod p_1 \bmod 2 \oplus S^4 \bmod p_2 \bmod 2 \\ &... \\ &... \end{aligned} \quad (5)$$

One may replace $p_1$ and $p_2$ by $n_1$ and $n_2$ that are products of primes. For better security, the primes should each be congruent to $3 \pmod 4$ as in the BBS generator [5].

## Recursive Random Number Generator

Here, we propose a recursive RNG based on Kak RNG which is motivated by the following goals:

1. Increasing the period of sum of d-sequence not only by a factor of $lcm\{(p_1 - 1), (p_2 - 1).......\}$ but by a multiple of it.

2. Smoothening of auto-correlation function.

The recursive formula proposed is as follows:

$$(S^i \bmod p_{11} + S^i \bmod p_{12} + \ldots S^i \bmod p_{1n})^k \bmod p_{21} \bmod 2 \quad \oplus$$
$$(S^i \bmod p_{11} + S^i \bmod p_{12} + \ldots S^i \bmod p_{1n})^k \bmod p_{22} \bmod 2 \quad \oplus \ldots$$
$$\ldots \quad (S^i \bmod p_{11} + S^i \bmod p_{12} + \ldots S^i \bmod p_{1n})^k \bmod p_{2m} \bmod 2 \qquad (6)$$

where $S$ is the seed and $p_{fg}$ is a prime number and $S$ and $p_{fg}$ are relatively prime to each other. The first subscript distinguishes the loops $i$ and $k$ and second subscript is number of that element in its respective loop.

It can be clearly seen from (6) that there are two loops to be traversed during the generation of random binary numbers. The outer loop is with respect to $k$ and the inner loop is with respect to $i$, i.e. loop $i$ is nested within the loop $k$.

**Algorithm:**

1. Seed $S$ is chosen to be a primitive element of all $p_{11}, p_{12}, \ldots, p_{1n}$.
2. Let $t = \max(i)$ be one period of $(S^i \bmod p_{11} + S^i \bmod p_{12} + \ldots S^i \bmod p_{1n})$.
3. Choose $p_{21}, p_{22}, \ldots p_{2m}$.
4. Choose an integer $u$ and let $\max(k) = u$. (*Note:* length of sequence $l = t \times u$.)
5. Execution of loop:
   a) Set $k = 1$.
   b) Generate random numbers by running $i$ from 1 to $t$.
   c) If $k \le u$, increment $k$ to $k+1$ and return to a).
   d) Else quit.

Before we derive the expression for the period of RNG, we define **SeedSet.**

***SeedSet:*** It represents one period (or a subset of one period) of random numbers generated by addition of d-sequences $S^i \bmod p_{11} + S^i \bmod p_{12} + \ldots S^i \bmod p_{1n}$. It is denoted as the set $S = \{ S_1, S_2, S_3, \ldots S_w \}$, where $w$ is a number the choice of which is a part of the design of the RNG.

**The length of the period**

If $S$ is a primitive element of $p_{11}, p_{12}, \ldots p_{1n}$, then the period of $S^i \bmod p_{11} + S^i \bmod p_{12} + \ldots S^i \bmod p_{1n}$ is $lcm\{ (p_{11}-1), (p_{12}-1), \ldots (p_{1n}-1) \}$.

If $S$ is a not a primitive element of $p_{11}, p_{12}, \ldots p_{1n}$, then, as mentioned earlier, the period of $S^i \bmod p_{11} + S^i \bmod p_{12} + \ldots S^i \bmod p_{1n}$ is a divisor of $lcm\{ (p_{11}-1), (p_{12}-1), \ldots (p_{1n}-1) \}$.

Let $P_s$ denotes the period of $(S^i \bmod p_{11} + S^i \bmod p_{12} + \ldots S^i \bmod p_{1n})$ and let the generated SeedSet be $S = \{ S_1, S_2, S_3, \ldots S_w \}$. We can write,

$$P_{p_{21}}1 = Period(S_1, p_{21}), \quad P_{p_{22}}1 = Period(S_1, p_{22}) \;\ldots\; P_{p_{2m}}1 = Period(S_1, p_{2m})$$
$$P_{p_{21}}2 = Period(S_2, p_{21}), \quad P_{p_{22}}2 = Period(S_2, p_{22}) \;\ldots\; P_{p_{2m}}2 = Period(S_2, p_{2m})$$

$$\cdots \qquad\qquad \cdots \qquad\qquad \cdots$$
$$P_{p_{21}}w = Period(S_w, p_{21}),\quad P_{p_{22}}w = Period(S_w, p_{22}) \ldots P_{p_{2m}}w = Period(S_w, p_{2m})$$

where $Period(S_q, p_{2r})$, denotes the period of d-sequence generated by seed $S_q$ with respect to $p_{2r}$ and $q$ varies from 1 to $w$ and $r$ varies from 1 to $m$.

Thus, the period of outer loop is,

$$\begin{aligned} P_{p_{21}p_{22}\cdots p_{2m}} &= lcm\,(\,P_{p_{21}}1, \ldots P_{p_{21}}w, \quad P_{p_{22}}1, \ldots P_{p_{22}}w, \quad \ldots \quad P_{p_{2m}}1, \ldots P_{p_{2m}}w) \\ &= lcm\,[\,lcm\,(P_{p_{21}}1, \ldots P_{p_{21}}w), \quad lcm\,(P_{p_{22}}1, \ldots P_{p_{22}}w), \quad \ldots \quad lcm\,(P_{p_{2m}}1, \ldots P_{p_{2m}}w)] \end{aligned}$$

The period of the output sequence is,

$$P = P_{p_{21}p_{22}\cdots p_{2m}} \times number\ of\ \ elements\ in\ the\ SeedSet\ (S) \qquad (7)$$ ■

If the inner loop is restricted to just one term, i.e. $p_{11}$, and $p_{21} < p_{11},\ p_{22} < p_{11},\ p_{23} < p_{11}, \ldots\ p_{2m} < p_{11}$, then the choice of seed $S$ as a primitive root of $p_{11}$ with a SeedSet consisting of one complete period will guarantee a maximum length sequence. This happens because the SeedSet generated will contain all the numbers less than $p_{11}$. At least one of the seeds from the SeedSet will be a primitive element of $p_{21},\ p_{22}, \ldots\ p_{2m}$, yielding a maximum period for all the primes.

We now consider a simple subset of (6) consisting only of two terms each in inner and outer loop and verify the expression for period (7). The subset that we consider is

$$(S^i \bmod p_{11} + S^i \bmod p_{12})^k \bmod p_{21} \bmod 2 \quad \oplus \quad (S^i \bmod p_{11} + S^i \bmod p_{12})^k \bmod p_{22} \bmod 2 \qquad (8)$$

**Example 1.** Let $S = 2,\ p_{11} = 3,\ p_{12} = 5,\ p_{21} = 7,\ p_{22} = 11.$ Since, 2 is a primitive element of $p_{11}$ and $p_{12}$, $P_s = lcm\,(2, 4) = 4$. SeedSet $S = \{\,4, 5, 5, 2\}$, and therefore,

| | | |
|---|---|---|
| $P_{p_{21}}1 = Period\ \ (4, 7) = 3$ | | $P_{p_{22}}1 = Period\,(4,11) = 5$ |
| $P_{p_{21}}2 = Period\ \ (5, 7) = 6$ | and | $P_{p_{22}}2 = Period\,(5,11) = 5$ |
| $P_{p_{21}}3 = Period\ \ (5, 7) = 6$ | | $P_{p_{22}}3 = Period\,(5,11) = 5$ |
| $P_{p_{21}}4 = Period\ \ (2, 7) = 3$ | | $P_{p_{22}}4 = Period\,(2,11) = 10$ |

Thus, $P_{p_{21}p_{22}} = lcm\,[\,lcm\,(3, 6, 6, 3), lcm(5, 5, 5, 10)\,] = lcm\,(6, 10) = 30$.

The final period of the output sequence is

$$P = P_{p_{21}p_{22}} \times \{number\ of\ \ elements\ in\ the\ seed - set\ (S)\} = 30 \times 4 = 120\ .$$

**Example 2.** Let $S = 2,\ p_{11} = 23,\ p_{12} = 29,\ p_{21} = 7,\ p_{22} = 11.$ It so turns out that 2 is a not a primitive element of either 23 or 29, and the period for the sum of the individual d-sequences is 308. The final period becomes $P = 30 \times 308 = 9240$.

## Results

We present results for the simple case (8). Since the auto-correlation function of a random sequence is two-valued, we wish to confirm that our sequences are close to this ideal. (*Note:* 1. Circular and linear auto-correlations are calculated by setting the 0s (zeros) in the binary output to -1, and the functions are not normalized. 2. The notation in the graphs: $p1 = p_{11}$, $p2 = p_{12}$, $p3 = p_{21}$, $p4 = p_{22}$.)

### Circular Auto-correlation

The graphs in Figure (1) show the effect of multiplication of periods. The seed $S = 2$ and primes $p_{21} = 23$ and $p_{22} = 29$ are kept constant. In the left graph, $p_{11}$ and $p_{12}$ are 3 and 7, respectively, and from (8) the period is $P = 1848$. In the right graph, $p_{11}$ and $p_{12}$ are 5 and 7, respectively, and from (8) the period is $P = 3639$.

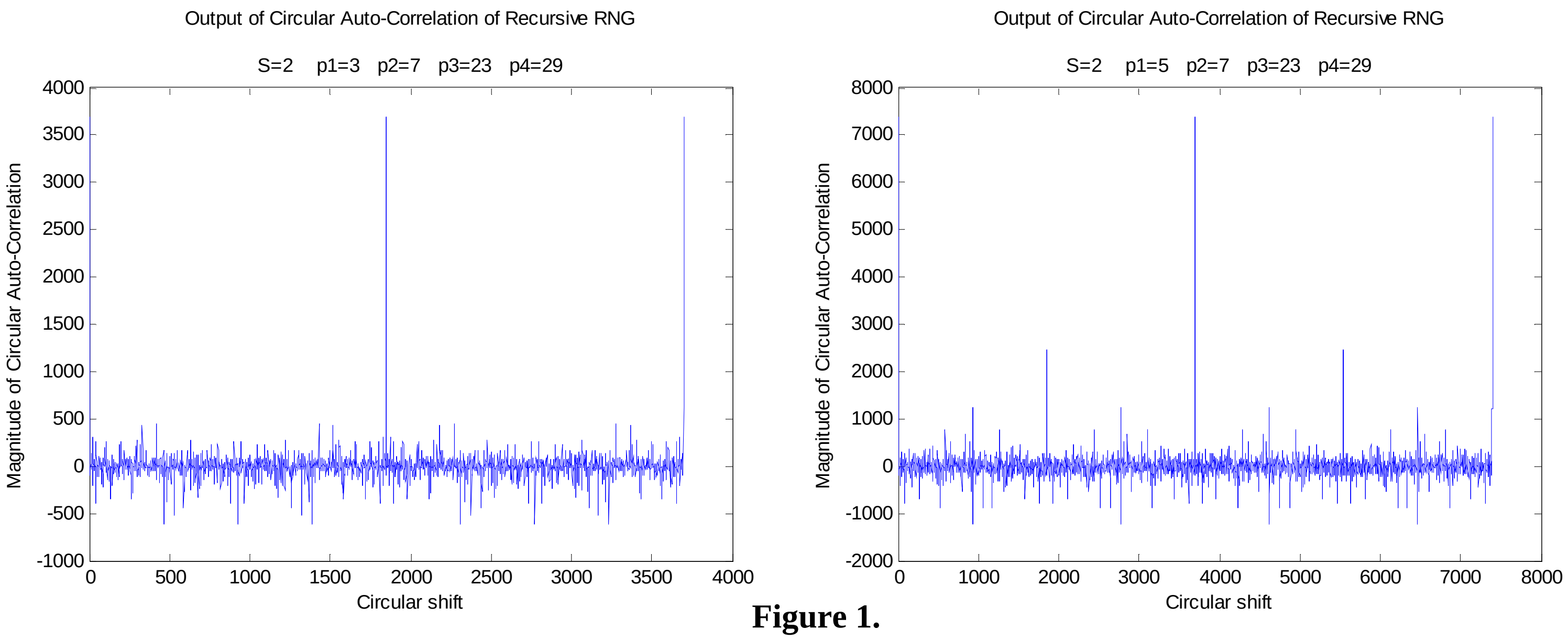

**Figure 1.**

The graphs in figure (2), illustrate the smoothening effect that Recursive RNG has on the auto-correlation functions of the output random numbers.

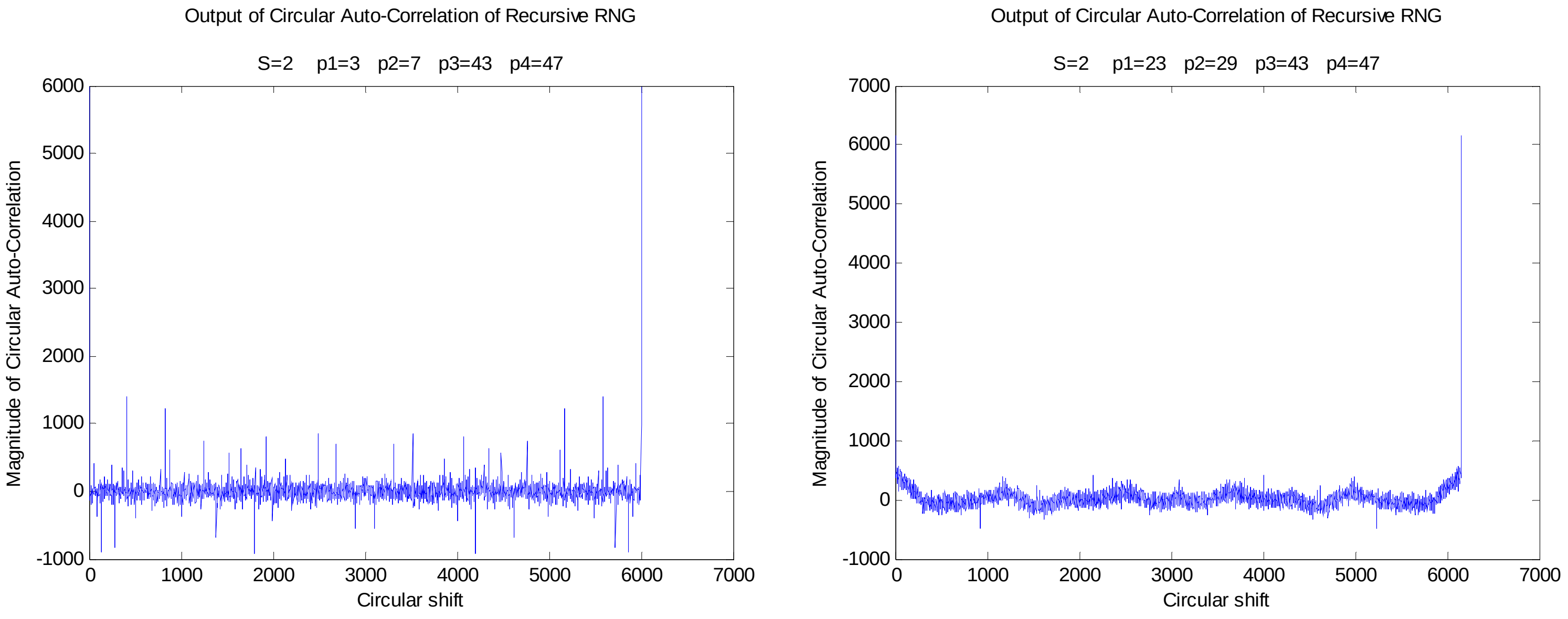

**Figure 2.**

It is thus observed that the choice of $p_{11}$ and $p_{12}$ are the factors that control the smoothening. Larger values of $p_{11}$ and $p_{12}$ give even better auto-correlation.

### Linear Autocorrelation

The linear auto-correlation of two experimental sequences is shown in figure (3). The comparison to be made in these graphs is that for sufficiently large prime numbers not only the period is very large but its linear autocorrelation approaches ideal approximation for RNG applications.

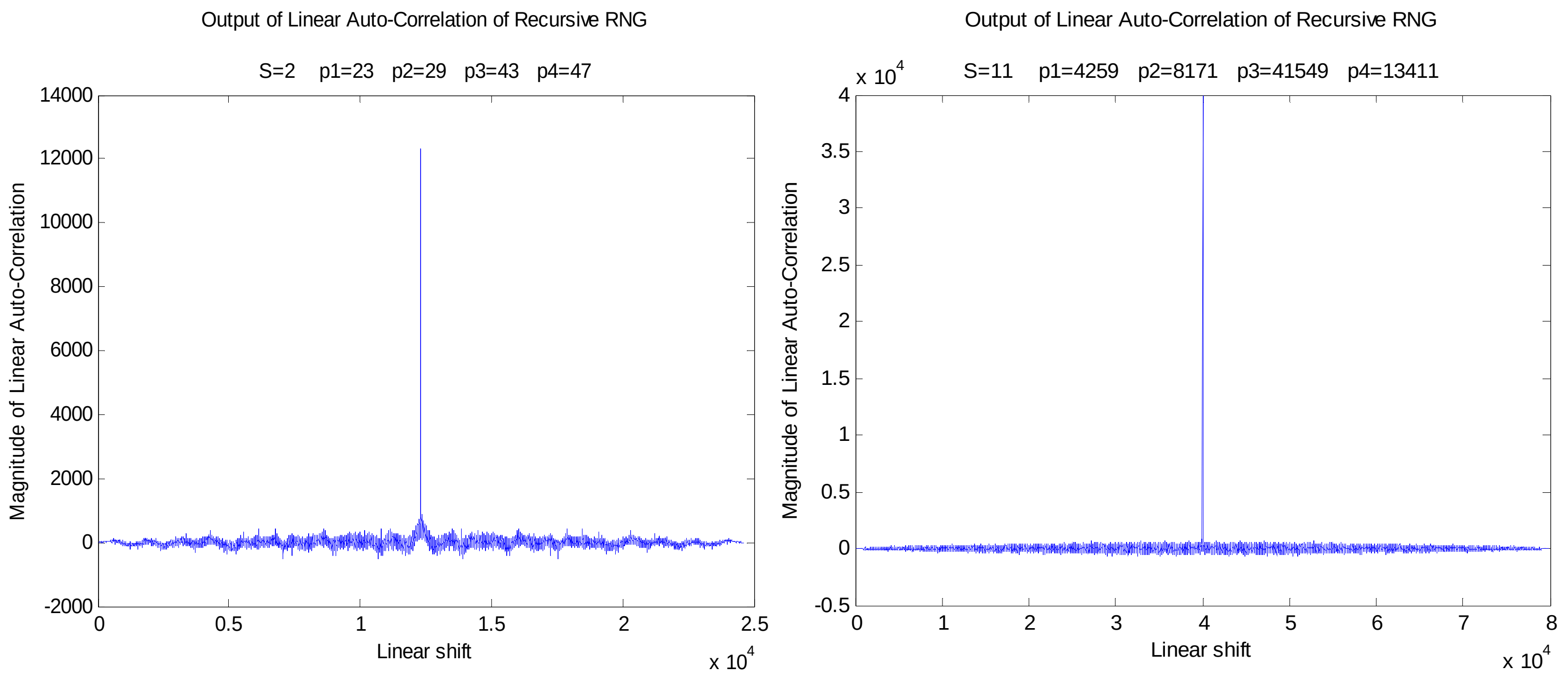

**Figure 3.**

Note that for small prime numbers one complete period has been taken into consideration. However, for large prime numbers, since the period is very large, only first 40,000 binary random numbers are taken into consideration.

## Conclusions

From the graphs above, it is clear that our goals of an RNG with a very large period and a good approximation to the ideal two-valued auto-correlation function were met. The period of the recursive RNG is in agreement to the theory.

The choice of $p_{11}$ and $p_{12}$ provides the user of the RNG flexibility to control the auto-correlation function to desired characteristics. The results above are for small prime numbers, but they hold good for large prime numbers as well. We have only presented results for a subset (6), and this can be extended to more than two d-sequences in order to obtain larger periods.

Recursive versions of other d-sequence based RNGs described in [4] may also be developed. Another extension would be to base a recursive generator on the cubic transformation [7].